\documentclass[twocolumn,twoside]{article}
\usepackage{imr}

\usepackage{amsmath,psfig,epsf,makeidx}
\usepackage{myconcrete,jeffe,euler,url}
\usepackage{amsfonts}
\usepackage[ruled]{algorithm}
\usepackage{algorithmic}
\usepackage{multirow}

\newcounter{theoctr}[part]

\newtheorem{definition}[theoctr]{Definition}
\newtheorem{theorem}[theoctr]{Theorem}

\newtheorem{lemma}[theoctr]{Lemma}

\def\Floor#1{\left\lfloor #1 \right\rfloor}
\def\Ceil#1{\left\lceil #1 \right\rceil}
\def\Seq#1{\left\langle #1 \right\rangle}
\def\Set#1{\left\{ #1 \right\}}

\def\Brack#1{\left[ #1 \right]}         
\def\IntervalCO#1{\left[ #1 \right)}         

\newenvironment{proof-idea}{\noindent{\bf Proof Idea}\hspace*{1em}}{\qed\bigskip}

\def\tt{{\mbox{\boldmath $t$}}}
\def\ss{{\mbox{\boldmath $s$}}}

\newcommand\BB{{\mbox{$\cal B$}}}

\newcommand\CC{{\mbox{$\cal C$}}}
\newcommand\DD{{\mbox{$\cal D$}}}
\newcommand\II{{\mbox{$\cal I$}}}
\newcommand\EE{{\mbox{$\cal E$}}}
\newcommand\GG{{\mbox{$\cal G$}}}

\newcommand\RR{{\mbox{$\Bbb R$}}}
\newcommand\ZZ{{\mbox{$\Bbb Z$}}}

\def\union{\cup}

\def\lfs{\mathrm{lfs}}

\newcommand{\comment}[1]{}

\begin{document}

\pagestyle{myheadings}
\markboth{Daniel A. Spielman, Shang-Hua Teng, and
        Alper \"Ung\"or}
        {Parallel Delaunay Refinement: Algorithms and Analyses}

\title{\uppercase{Parallel Delaunay Refinement:\\ 
      Algorithms and Analyses}}
\author{Daniel A. Spielman$^1$ \and Shang-Hua Teng$^2$ \and 
	Alper \"Ung\"or$^3$}

\date{
$^1$Department of Mathematics, Massachusetts Institute of Technology, 
  spielman@math.mit.edu\\
$^2$Department of Computer Science, Boston University 
  and Akamai Technologies Inc.,  steng@cs.bu.edu\\
$^3$ Department of Computer Science, 
  Duke University, ungor@cs.duke.edu}

\abstract{
In this paper, we analyze the complexity
  of natural parallelizations of Delaunay refinement
  methods for mesh generation.
The parallelizations employ a simple strategy:
  at each iteration, they choose a set of ``independent''
  points to insert into the domain, and then update the Delaunay
  triangulation.
We show that such a set of independent points can be constructed
  efficiently in parallel and
  that the number of iterations needed is $O(\log^2(L/s))$,
  where $L$ is the diameter of the domain, and $s$ is the
  smallest edge in the output mesh.
In addition, we show that the insertion of 
  each independent set of points
  can be realized sequentially by Ruppert's method in two dimensions
  and Shewchuk's in three dimensions.
Therefore, our parallel Delaunay refinement methods
  provide the same element quality and mesh size guarantees 
  as the sequential algorithms in both two and three dimensions.   
For quasi-uniform meshes, such as those produced by Chew's method, we show that
  the number of iterations can be reduced to $O(\log(L/s))$.
To the best of our knowledge,
  these are the first provably polylog$(L/s)$ parallel time Delaunay
  meshing algorithms that generate well-shaped meshes of size optimal
  to within a constant.
}

\maketitle

\section{Introduction}

\setlength{\baselineskip}{1.\baselineskip}

Delaunay refinement is a popular and  practical technique for
  generating  well-shaped unstructured 
  meshes \cite{LiTeng01,Ruppert93,Shewchuk98}.
The first step of a Delaunay refinement algorithm is 
  the construction of
  a constrained or conforming Delaunay triangulation of 
  the input domain.
This initial Delaunay triangulation need not be well-shaped.
Delaunay refinement 
  then iteratively adds new points to the domain to improve the
  quality of the mesh and to make 
  the mesh respect the boundary of the input domain.
A sequential Delaunay refinement algorithm typically adds
  one new vertex per iteration, although sometimes one may prefer to insert
  more than one new vertex at each iteration.
Each new point or set of points is
  chosen from a set of potential candidates ---
  the circumcenters of poorly conditioned simplices (to improve mesh
  quality) and the diameter-centers of boundary simplices (to 
  conform to the domain boundary).
Ruppert \cite{Ruppert93} was the first to show that the
  proper application of Delaunay refinement produces
  well-shaped meshes in two dimensions whose size is within a small
  constant factor of the best possible.
Ruppert's result was then extended to three dimensions 
  by Shewchuk \cite{Shewchuk98} and Li and Teng \cite{LiTeng01}.
Efficient sequential Delaunay refinement software has been
  developed both in two \cite{Ruppert93,Shewchuk96} and
  three dimensions \cite{Shewchuk98}.
Chrisochiedes and Nave \cite{ChrisN99}
and Okusanya and Peraire \cite{OkusanyaP96} developed parallel
  software using Delaunay refinement, 
  for which they have reported good performance.
Recently, Nave {\em et al.}\ \cite{NaveCC02} 
  presented a parallel Delaunay refinement algorithm and proved
  that it produces well-shaped meshes.
The complexity of their algorithm as well as the size of the mesh 
  it outputs remains unanalyzed.

In this paper, we study the parallel complexity of a 
  natural parallelization
  of Delaunay refinement.
One of the main ingredients of our parallel method
  is a notion of independence among
  potential candidates for Delaunay insertion at each iteration.
Our parallel Delaunay method performs the following steps 
  during each iteration.
\vspace{-6pt}
\begin{enumerate}
\setlength{\itemsep}{-2pt}
\item Generate an independent set of points for parallel insertion;
\item Update the Delaunay triangulation in parallel.
\end{enumerate}
\vspace{-6pt}

Our independent sets have the following properties:
\vspace{-6pt}
\begin{itemize}
\setlength{\itemsep}{-2pt}
\item Their insertion 
  can be realized sequentially by Ruppert's method in 2D
  and Shewchuk's in 3D.
Hence, an algorithm that inserts all their points in parallel
  will inherit the guarantees of Ruppert's and Shewchuk's methods
  that the output mesh be well-shaped and have size optimal up to a constant.
\item The independent sets can be
  generated efficiently in parallel. In addition,
  they are ``large enough'' so that the
   number of parallel iterations needed is $O(\log^2(L/s))$,
   where $L$ and $s$ are the diameter of the domain and the
  smallest edge in the output mesh, respectively.
\item When a quasi-uniform mesh is desired as in Chew's method, 
  the number of iterations can be reduced to $O(\log(L/s))$.
\end{itemize}
\vspace{-6pt}
We should emphasize here that our analysis focuses on the number of
  parallel iterations of Delaunay refinement. 
The independence of the new points 
   do not necessarily 
  imply a straightforward parallel insertion scheme at each iteration. 
There are several existing parallel Delaunay 
  triangulation algorithms that we can employ at each iteration.
For example, in 2D we can use the divide-and-conquer parallel
  algorithm developed by Blelloch {\em et al.}\ 
  \cite{BlellochHMT99} for Delaunay triangulation. 
Their algorithm uses $O(n \log n)$ work and $O(\log^3 n)$ parallel
  time. 
We can alternatively use the randomized parallel algorithms of Reif and Sen 
 \cite{ReifS99}, or by Amato {\em et al.}\ \cite{AmatoGR94}, in both
  two and three dimensions.
Both of these randomized parallel Delaunay triangulation algorithms
  have expected parallel running time $O (\log n)$.
Using one of these adds a logarithmic factor to our worst-case 
  total parallel time complexity analysis.
To the best of our knowledge,
  these are the first provably polylog$(L/s)$ parallel time Delaunay
  meshing algorithms that generate well-shaped meshes of size optimal
  to within a constant.

\subsection{Motivation and Related Work}

This work is motivated by the observation that both sequential and
  parallel implementations of  Delaunay refinement algorithms
  seem to produce the best meshes in practice.
However, improvements in the speed of parallel numerical solvers
  are creating the need for comparable speedups in meshing software:
{L\"ohner} and Cebral \cite{LohnerC99} have reported that
  improvements in parallel numerical solvers \cite{ToppingK95}
  have resulted in the simulation time of numerous
  practical systems being dominated by the meshing process.

Quadtree-based methods are an alternative to Delaunay refinement. 
They also generate well-shaped meshes whose size is within a constant 
  factor of the best possible \cite{BernEG94,MitchellV92}.
In practice, however, they often generate meshes larger than
   Delaunay refinement on the same input.
The parallel complexity of the 
   quadtree-based methods is nevertheless better understood.

Several parallel mesh generation algorithms have been developed.
On the theoretical extreme, Bern, Eppstein and Teng \cite{BernET99}
  gave a parallel $O(\log n)$ time algorithm  using  $K/\log n$
  processors to compute a well-shaped quadtree mesh, where
  $K$ is the final mesh size.
There is also a simple level-by-level quadtree-based method that 
  is used in practice \cite{ShephardFBCOS97,SpielmanT02}.
One can easily show that this level-by-level based
  method takes $O (\log (L/s)+ K/p)$ parallel time,
  using $p$ processors \cite{SpielmanT02}.

Building upon \cite{BernET99}, 
  Miller {\em et al.}\ \cite{MillerTTW95} developed
  a parallel sphere-packing based Delaunay
  meshing algorithm that generates well-shaped
  Delaunay meshes of optimal size in 
  $O (\log n)$ parallel time using $K/\log n$ processors.
Their method uses a parallel maximal independent set algorithm \cite{Luby86}
  to directly generate the set of final mesh points, and
  then constructs the Delaunay mesh using parallel Delaunay triangulation.
As this algorithm has not been implemented, we do not know how the
  meshes it produces will compare.

Various parallel Delaunay refinement methods have been implemented
  and been seen to have good performance
  \cite{ChrisN99,LiTeng98,LohnerC99,OkusanyaP96}. 
These methods address some important issues such as 
  how to partition the domain so as
  to minimize the communication cost among the 
  processors.
Our new analysis on the number of parallel iterations of Delaunay
  refinement could potentially provide provable
  bounds on their parallel complexity.

Our work also helps explain the performance of
  some sequential implementations of Delaunay refinement, 
  especially those which use a 
  Delaunay triangulator as a black-box.
In such situations, it is often desirable
  to minimize the number of calls to the black-box Delaunay triangulator
  by inserting multiple points at each iteration.
Our bounds on the number of iterations provide a bound on
  the number of calls to the Delaunay triangulator.

We omit the proofs of Lemmas
\ref{lem:rupertConservation},
\ref{lem:upgrade},
\ref{lem:preprocessing},
\ref{lem:techruppert}, and
\ref{lem:lfs_ratio}
 and Theorems 
\ref{thm:sequential_parallel},
\ref{thm:parallelchew},
\ref{thm:sequential_parallel_periodic_3D}, and 
\ref{thm:parallelShewchuk}
in this version due to page limitation.
A full version of the paper is available at
\url{http://www.cs.duke.edu/~ungor/abstracts/parallelDelRef.html}.

\section{Preliminaries} \label{sec:pre}
\subsection{Input Domain}
\label{sec:input}

In 2D, the input domain $\Omega $ is represented as a {\em planar
  straight line graph}\index{planar straight line graph (PSLG)}
  (PSLG) \cite{Ruppert93} --- a proper planar drawing
  in which each edge is mapped to
  a straight line segment between its two endpoints.
The segments express the {\em boundaries} of $\Omega $
  and the endpoints are the {\em vertices} of $\Omega $.
The vertices and boundary segments of $\Omega $ will be
  referred to as {\em input features} of $\Omega$.
A vertex is incident to a segment if it is one of the endpoints of the
  segment.
Two segments are incident if they share a common vertex.
In general, if the domain is given as a collection of vertices only, then 
 the boundary of its convex hull is taken to be the
 boundary of the input.

Miller {\em et al.}\ \cite{MillerTTWW96} presented a natural
  extension of PSLGs, called {\em piecewise linear complexes} (PLCs),
  to describe domains in three and higher dimensions.
In three dimensions, the domain $\Omega$ is a collection
  of vertices, segments, and facets where
  (i) all lower dimensional elements on the boundary of an element
  in $\Omega$ also belongs to $\Omega$, and
  (ii) if any two elements intersect, then their intersection is a
  lower dimensional element in $\Omega$.
In other words, a PLC in $d$ dimensions is a cell complex
  with polyhedral cells from $0$ to $d$ dimensions.

\subsection{Delaunay Triangulation}

Let $P$ be a point set in $\RR^d$. 
A simplex $\tau$ formed by a subset of $P$ points is a 
  {\em Delaunay simplex} \index{Delaunay simplex} 
  if there exists a circumsphere of $\tau $ whose interior
  does not contain any points in $P$.
The {Delaunay triangulation}  \index{Delaunay triangulation}
  of $P$, denoted $Del(P)$, is a PLC that contains all Delaunay
  simplices. 
If the points are in general position, 
  that is, if no $d+2$ points in $P$ are co-spherical, 
  then $Del(P)$ is a simplicial complex. 

The Delaunay triangulation of a point set can be constructed
  in $O (n\log n)$ time in 2D
  \cite{ClarksonS89,GuibasKS90,Fortune92}
  and in $O(n^{\lceil d/2 \rceil})$ time in $d$ dimensions 
  \cite{ClarksonS89, Seidel92}.
A nice survey of these algorithms can be found in 
  \cite{Fortune92}.

One way to obtain a triangulation that conforms
  to the boundary of a PSLG domain is to use
  a {\em Constrained Delaunay triangulation}. 
Let $P$ be the set of vertices of a PSLG $\Omega$.
Two points $p$ and $q$ in $P$ are said to be
  visible from each other
  if the line segment $pq$ does not intersect the interior 
  of any segment in $\Omega$. 
Three points form a constrained Delaunay triangle
  if the interior of their circumcircle contains no point from $P$
  that is visible from all three points.
The union of all constrained Delaunay triangles forms a
  {\em constrained Delaunay triangulation} $CDT(\Omega)$.
  \index{constrained Delaunay triangulation}
Chew developed an algorithm for 
  computing constrained Delaunay
  triangulations \cite{Chew89b}.

A Delaunay triangulation $T$ of input and Steiner points 
  is a {\em conforming Delaunay triangulation} of a PLC $\Omega $
  \index{ conforming Delaunay triangulation}
  if every face of $\Omega$ is a union of faces of $T$.
In 2D, Edelsbrunner and Tan proved that $O(n^3)$
  additional points are sufficient to generate a conforming triangulation 
  of a PSLG of complexity $n$ \cite{EdelsbrunnerT93}.
A 2D solution proposed by Saalfeld \cite{Saalfeld91} 
  is extended to 3D by Murphy {\em et al.}\ \cite{MurphyMG00} and 
  Cohen-Steiner {\em et al.}\ \cite{CohenVY02}. 
However, it remains open whether the size of their output
  is polynomial in the input size or local feature size.
The definition of local feature size will be given in Section 
  \ref{sec:DelaunayRefinement}.
When the angle between the
  faces of a PLC is bounded from below, say for example
  by $\pi /2$, then one can apply Delaunay refinement 
  to generate well-shaped conforming triangulations 
  whose size is close to optimal 
  both in two \cite{Chew89, Ruppert93} 
  and three dimensions \cite{Chew97, LiTeng01, Shewchuk98}. 

\section{2D Sequential Delaunay Refinement}
\label{sec:DelaunayRefinement}

In this section, we recall Ruppert's and
  Chew's algorithms for constructing Delaunay
  meshes of PLSGs in 2D.
Following Ruppert \cite{Ruppert93}, 
  we assume that the angle between two adjacent 
  input segments is at least $\pi /2$.
Boundary treatments that relax this assumption are discussed in 
  \cite{Ruppert93, Shewchuk02}.

In the process of Delaunay refinement, one could either
  maintain a constrained Delaunay triangulation, or one 
  just keeps track of the set of input segments that are not respected.
The first approach does not extend to three dimensions because,
  in 3D, some PLCs do not have a constrained Delaunay triangulation. 
We therefore use the second approach. 

At each iteration, we choose a new point for insertion from a set of
  candidate points.
There are two kinds of candidate points: (1) the circumcenters of
  existing triangles, and (2) the midpoints of existing boundary
  segments.

\begin{figure}
\begin{center}
\begin{tabular}{c}
\psfig{figure=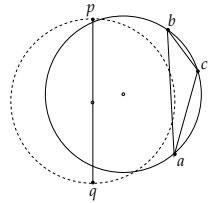,height=.16\textwidth} 
\end{tabular}
\caption{Circumcenter of  triangle $abc$ encroaches the segment $pq$.}
\label{fig:candidates}
\end{center}
\end{figure}

Let the diametral circle of a segment be the circle 
 whose diameter is the segment.
A point is said to {\em encroach} a segment if it is 
  inside the segment's diametral circle. 
  (See Figure \ref{fig:candidates}.)

At iteration $i$, the circumcenter of a triangle is a 
  {\em potential candidate} for insertion if
  the triangle is {\em poorly shaped}.
For example, in Ruppert's algorithm, a triangle 
  is considered poorly shaped if the ratio of
  its circumradius to the length of its shortest side is larger
  than a pre-specified constant $\beta_{R} \geq \sqrt{2}$.
Let $\dot{\CC}^{(i)}$ denote
  the set of all potential candidate circumcenters 
  that do not encroach any segment.
Let $\CC^{(i)}$ denote their corresponding
  circumcircles.
Similarly, let $\dot{\BB}^{(i)}$ denote the set of all potential candidate circumcenters 
  that do encroach some segment.
Let $\BB^{(i)} $ denote their corresponding circumcircles.

The midpoint of a boundary segment
  is a {\em candidate} for insertion if (1) the segment is not part of
  the current Delaunay triangulation, that is, its diametral circle is
  encroached by some existing mesh points, or (2) its segment
  is encroached by a circumcenter in $\dot{\BB} $.
In the latter case, this potential circumcenter candidate is {\em rejected}
  from insertion.
Let $\dot{\DD}^{(i)}_{T}$ be all midpoint candidates of type (1) and let   
 $\dot{\DD}^{(i)}_{\BB}$ be all midpoint candidates of type (2).

\begin{algorithm}[h]
\caption{Sequential Delaunay Refinement} 
\label{alg:sequential}
\begin{algorithmic}
\REQUIRE A PSLG domain $\Omega$ in $\RR^2$
\STATE Let $T$ be the Delaunay triangulation 
      of the vertices of $\Omega$. Let $i = 0$ and compute 
      $\BB^{(i)}$, $\CC^{(i)}$,  $\DD^{(i)}_{T}$,  and $\DD^{(i)}_{\BB }$;
\WHILE{$\CC^{(i)} \union \DD^{(i)}_{T} \cup \BB^{(i)}$ is
not empty}
\STATE Choose a point $q$ from $\dot{\CC}^{(i)} \union
\dot{\DD}^{(i)}_{T} \cup \dot{\DD}^{(i)}_{\BB}$ and insert $q$ into the triangulation. 
       If $q$ is a midpoint of a
       segment $s$, remove $s$ from the segment list and replace it
       with two segments from $q$ to each endpoint of $s$;
\STATE Update the Delaunay triangulation $T$; $i=i+1$;
\STATE Compute $\dot{\BB}^{(i)}$, $\dot{\CC}^{(i)}$,  $\dot{\DD}^{(i)}_{T}$,  and $\dot{\DD}^{(i)}_{\BB}$.
\ENDWHILE
\end{algorithmic}
\end{algorithm}

The points inserted by the Delaunay refinement are often called
  {\em Steiner points}.

If a quasi-uniform mesh,
  such as that produced by Chew's method, is desired \cite{Chew89},  
  then we use the following notion of poorly shaped triangle:
  A triangle is {\em poorly-shaped} if the ratio of its circumradius 
  to the length of the shortest edge in the current Delaunay
  triangulation $T$ is more than a pre-specified constant 
  $\beta_{C} \ge \sqrt{2}$.

\begin{figure}[bth]
\begin{center}
\begin{tabular}{c@{\ }c}
\psfig{figure=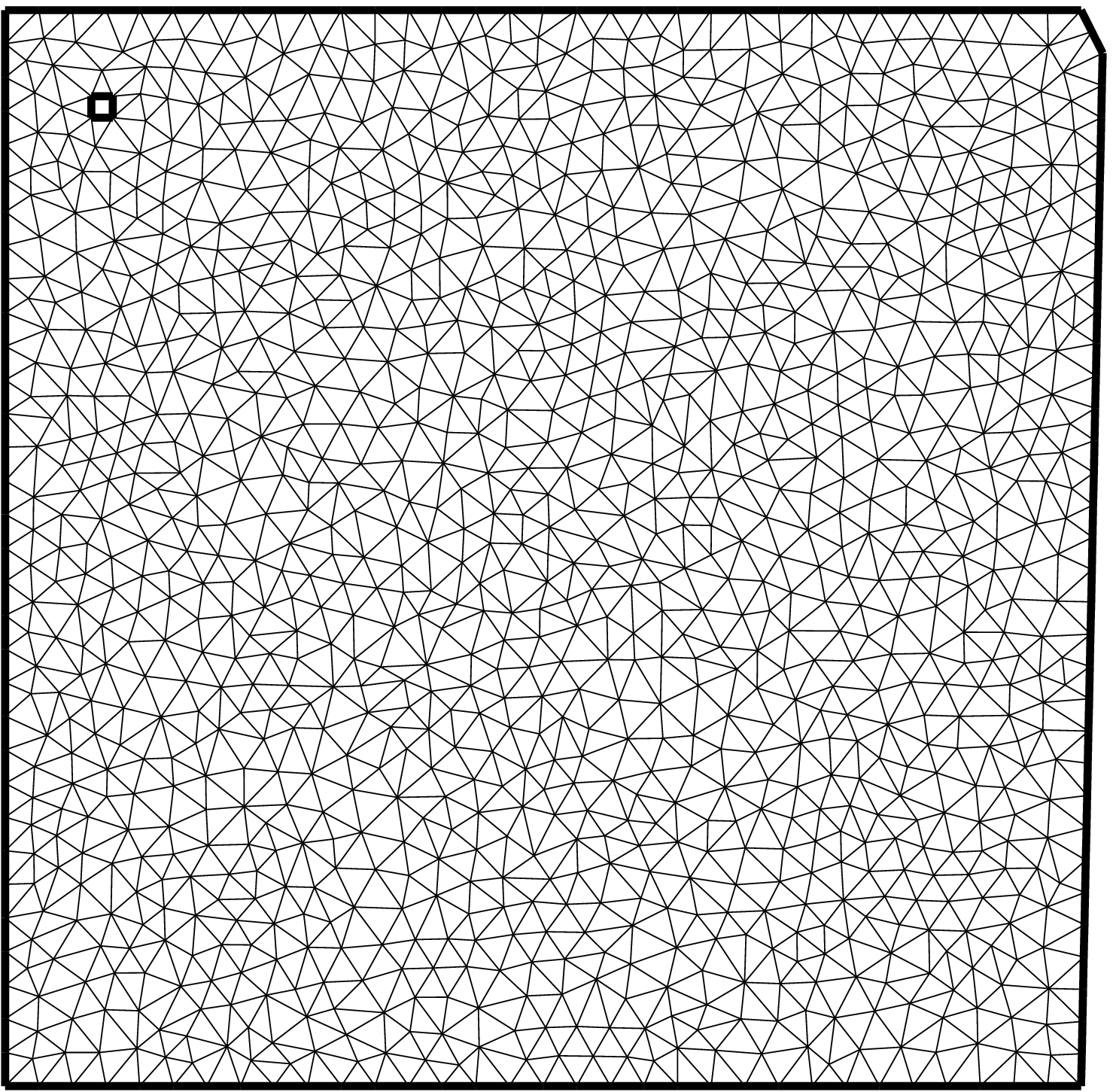,height=1.4in} &
\psfig{figure=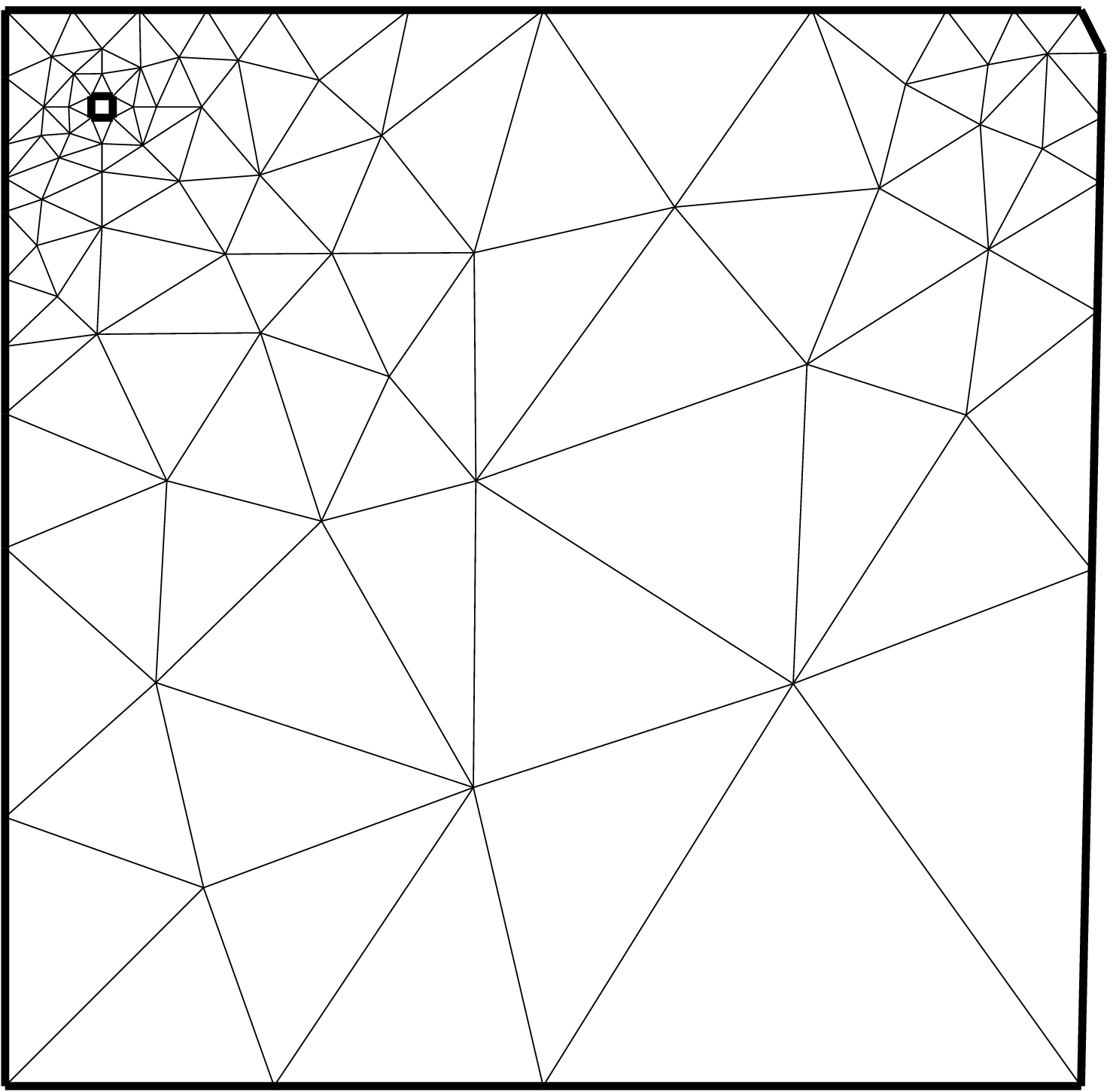,height=1.4in} \\
(a) & (b) \\
\end{tabular}
\end{center}
\caption{
{The output of (a) Chew's and (b) Ruppert's algorithm} on the
  same input. Both of these meshes have minimum angle $ > 29^\circ$. 
  The first mesh has $2246$ and the second has $131$ elements.}
\label{fig:ChewRuppert}
\end{figure}

Figure \ref{fig:ChewRuppert} shows the output of the Delaunay
  refinement illustrating the difference between Chew's 
  and Ruppert's refinement.  
We call these two variants of the Delaunay refinement algorithm 
  {\em Chew's algorithm} and {\em Ruppert's algorithm}.

In their original papers \cite{Chew89,Ruppert93}, 
  Chew and Ruppert presented their Delaunay refinement algorithms
  as particular variations of Algorithm \ref{alg:sequential} ---they specified 
  how to choose the next point at each iteration
  from the set of candidates.
In this paper, we will consider the following variation 
  of Algorithm \ref{alg:sequential} which is more 
  aggressive in adding boundary points --- 
we choose this variation to parallelize because its analysis
  is relatively simpler to present.

In this variation, $\BB^{(i)} $, $\CC^{(i)} $, and $\DD^{(i)}_{T}$ are the same
  as in Algorithm \ref{alg:sequential}.
The set $\DD^{(i)}_{\BB}$ is built incrementally.
At iteration $i$, we compute $\BB^{(i)} $ first and 
   let $\DD^{(i)}$ 
   be the set of diametral circles that are encroached by some circumcenters
   of $\BB^{(i)} $.
We then set $\DD^{(i)}_{\BB } = \DD^{(i-1)}_{\BB } \cup  \DD^{(i)} $.

In other words, if a segment is encroached by a circumcenter of a
  poorly-shaped Delaunay triangle, its midpoint will be added to
  the set of candidate midpoints and remains candidate thereafter.
This is in contrast with Algorithm \ref{alg:sequential}, 
  in which an encroached midpoint
  is added to the set of candidate midpoints only for the next
  iteration.
If another candidate is chosen that is in a circumcircle whose
  center encroaches the segment, the circumcircle will no longer be in
  the Delaunay triangulation at the end of the iteration, and
  hence the segment might not be encroached in the triangulation at
  the end of the iteration.
So, its midpoint might not be a candidate in future iterations.

Assuming that the angle between two adjacent 
  input segments is at least $\pi /2$,
Chew's algorithm terminates with well-shaped quasi-uniform meshes,
  while 
  Ruppert's algorithm \cite{Ruppert93} terminates with a well-shaped Delaunay 
  mesh of the input domain whose elements adapt to the local geometry
  of the domain.
The number of triangles in the mesh generated by Ruppert's algorithm 
 is asymptotically optimal up to a constant.
The proofs of Ruppert's and Chew's \cite{Chew89,Ruppert93} that their
  algorithms terminate with a well-shaped mesh of size within a
  constant factor of optimal can be easily extended to our variation
  of Algorithm \ref{alg:sequential} discussed above.
We refer interested readers to \cite{Ruppert93} and \cite{Shewchuk02}.
Here we give a high level argument and introduce an important concept
that will be used in Section \ref{sec:preprocessing} for
 preprocessing an input domain in parallel.

Given a domain $\Omega $, the {\em local feature size} of each point
   $x$ in $\Omega $, denoted by $\lfs_{\Omega} (x)$, is the radius
   of the smallest disk centered at $x$ that touches two non-incident
   input features.
Ruppert showed that
  every Delaunay triangle in the final mesh is well-shaped
  and that the length of the longest edge in each Delaunay triangle
  is within a constant factor of $\lfs_{\Omega} (x)$ for each
  $x$ in the interior of the triangle.

Suppose $M$ is a mesh generated by our 
   variation of Algorithm \ref{alg:sequential}.
Let $\Omega '$ be the domain obtained from $\Omega $ by adding to
  $\Omega $ all  mesh points in $M$ that are on the boundary segments
  of $\Omega $.
Then we can show (i) for all $x$ in $\Omega$, $\lfs_{\Omega} (x)$
  and $\lfs_{\Omega'} (x)$ are within a small constant factor of each
  other; and (ii) $M$ can be obtained by applying Ruppert's (or Chew's)
  variations of Algorithm \ref{alg:sequential} to $\Omega '$.
Therefore, the mesh produced by our variation of 
  Algorithm \ref{alg:sequential} has size within a small constant 
  factor of the
  one generated by Ruppert's (or Chew's) refinement method.

\section{Parallel 2D Delaunay Refinement}
\label{sec:parallelDelaunayRefinement}

To better illustrate our analysis of parallel
  Delaunay refinement, we first 
  focus on the case
  in which the input is a periodic point set (PPS) as introduced by
  Cheng {\em et al.}\ \cite{ChengDEFT99}. 
See also \cite{Edelsbrunner01}.
We will then extend our results to 
  produce boundary conforming meshes 
  when the input domain is a PSLG.

\subsection{Input Domain: Periodic Point Sets}
\label{sec:periodic}

If $P$ is a finite set of points in the half open unit square 
  $[0,1)^2$ and $\ZZ^2$ is the two dimensional integer grid,
  then  $S= P + \ZZ^2$ is a periodic point set
  \cite{Edelsbrunner01}.
The periodic set $S$ contains all points $p+v$, 
  where $p \in P$ and $v$ is an integer vector.
The Delaunay triangulation of a periodic point set is also periodic.

As $P$ is contained in the unit square, the diameter
  of $P$ is $L \leq \sqrt{2}$.
When we refer to the diameter of a periodic point set,
  we will mean the diameter of $P$.

\subsubsection{A generic parallel algorithm (PPS)}
\label{sec:genericParallelDelaunayRefinement}

For a periodic point set, the only candidates for 
  insertion are the circumcenters of poorly shaped triangles.
We need a rule for choosing a large subset of the candidates
  with the property that a sequential Delaunay refinement
  algorithm would insert each of the points in the subset.
Our rule is derived from the following 
  notion of {\em independence} among candidates.

\begin{definition}[Independence]
\label{def:conflict}
Two circumcenters $\dot{c}$ and $\dot{c}'$
  (and also the corresponding circles $c$ and $c'$)
  are {\em conflicting} if both $c$ and $c'$ contain
  each other's center.
Otherwise, $\dot{c}$ and $\dot{c}'$ (respectively $c$ and $c'$) are said to be 
  {\em independent}. 
\end{definition}

If two candidates conflict, at most one of them can be inserted.
Our rule is to insert a maximal independent set (MIS) of candidates
  at each iteration.
We will show that if an algorithm follows this rule, then it
  will terminate after a polylogarithmic number of rounds.

\begin{algorithm}
\caption{Generic Parallel Delaunay Refinement} 
\label{alg:genericParallelDelaunayRefinement}
\begin{algorithmic}
\REQUIRE A periodic point set $P$ in $\RR^2$

\STATE Let $T$ be the Delaunay triangulation 
      of $P$ 
\STATE Compute  $\dot{\CC}$, the set of all candidate circumcenters
  in $T$
\WHILE{$\dot{\CC}$ is not empty}
\STATE Let $\II$ be an independent subset of $\dot{\CC}$
\STATE Insert all the points in $\II$ in parallel 
\STATE Update $T$ and $\dot{\CC}$
\ENDWHILE
\end{algorithmic}
\end{algorithm}

In the next few subsections, we will discuss
  how to generate the independent sets used
  by the algorithm.
But first, we prove that regardless of how one chooses the independent
  set, our parallel algorithm can be sequentialized.
This implies that the algorithm inherits the guarantee of its
  sequential counterpart that it generates a 
 well-shaped mesh of size that is within a
  constant factor of optimal.

\begin{theorem}
\label{thm:sequential_parallel_periodic}
Suppose $M$ is a mesh produced by an execution 
  of the Generic Parallel Delaunay Refinement algorithm.
Then $M$ can be obtained by some execution of 
  one of the sequential Delaunay refinement
  algorithms discussed in Section \ref{sec:DelaunayRefinement}.
\end{theorem}

\begin{proof}
Let $\II_1, \II_2, \hdots, \II_k$ be the sets of vertices inserted 
  by the parallel algorithm above at iterations 
  $1, \hdots, k$, respectively.
We describe a sequential execution that inserts 
  all the points in $\II_i$ before any point of 
  $\II_j$ for $i<j$.
For each independent set $\II_i$, we insert
  the candidates according to their circumradius in the order from
  largest to smallest.
For any two circumcenters $\dot{a},\dot{b} \in \II_i$,
  assume that the radius of $a$ is larger than the radius of 
  $b$.
This implies that $\dot{a}$ can not be in the 
  circumcircle of $\dot{b}$, because $\dot{a}$ and $\dot{b}$ are
  independent.
Therefore, the insertion of $\dot{a}$ will not eliminate the triangle
  of $\dot{b}$.

Furthermore, observe that in any sequential execution, 
  the insertion of
  point $\dot{p} \in \II_i$ can not eliminate the triangle
  corresponding to 
 $\dot{q} \in \II_j$ for any $i < j$, for otherwise, $\dot{q}$ 
  would not exist in the $j^{th}$ iteration of the parallel execution. 

Therefore, the parallel and sequential executions terminate with the 
  same Delaunay mesh.
\end{proof}

To minimize the number of iterations, intuitively, we should
  choose a maximal independent set of candidates 
  at each iteration.
In Section \ref{sec:MIS}, we will give
   a geometric algorithm that computes
   a maximal independent set of candidates 
   efficiently in parallel. 
Our algorithm makes use of the following observation.

\begin{lemma}\label{lem:conflict_reject_periodic}
Suppose $c_{a}$ and $c_{b}$ are two conflicting circumcircles
   at iteration $i$,
  and let $r_{a}$ and $r_{b}$ be their circumradii.
Then $r_b/2 < r_a < 2r_b$.

\end{lemma}

\subsubsection{Parallelizing Chew's Refinement (PPS)}
\label{sec:chew_periodic}

In this section, we show that our parallel implementation 
  of Chew's refinement only needs $O (\log (L/s))$ iterations.
The basic argument is very simple --- we will show that
  the radius of the largest Delaunay circle reduces by a factor of 3/4 
  after some constant number (e.g., 98) of iterations.
Because the largest circumradius initially is $O (L)$ and 
  the largest circumradius in the final mesh is $\Omega (s)$,
  the iteration bound of $O (\log (L/s))$ follows immediately.

\begin{lemma}
\label{lem:chew2d_constant_iterations_periodic}
For all $i$, let $r_i$ be the largest circumradius of 
  a triangle in the
  Delaunay triangulation at the end of iteration $i$. 
For all $k\geq 98$, $r_{k} \le 3r_{k-98}/4$.
\end{lemma}
\begin{proof}
We assume by way of contradiction that
$r_{k} > 3 r_{k-98}/4$.
Let $i = k-98$. 
Let $c_k$ be a circumcircle with radius $r_{k}$
  after iteration $k$.
Let $\dot{c}_k$ be the center of $c_{k}$.

For $j \leq k$, it is clear that $c_{k}$ 
  is also an empty circle in iteration $j$, because
  the refinement process only adds new points.
But, $c_{k}$ might not be a circumcircle at iteration $j$.
We now show that for each iteration $j$, where $ i \leq j \le k$,
  there exists a circumcircle $c'_j$ with center $\dot{c}'_j$,
  and radius $r'_j$ such that
(1) $ ||\dot{c}'_j -\dot{c}_{k}|| \le 3r_i/4$ and 
(2) $r'_j \geq 3r_i/4$. 
  
Let $p_{k}$, $q_{k}$ and $t_{k}$ be 
  the vertices of the Delaunay triangle
  at iteration $k$ that defines $c_{k}$.
We will alter  $c_{k}$ in three stages
  to produce a suitable $c'_{j}$ that is the circumcircle of
  three points that exist at stage $j$: $p_j$, $q_j$ and $t_j$.

\vspace{-6pt}
\begin{enumerate}
\setlength{\itemsep}{-2pt}
\renewcommand{\theenumi}{\roman{enumi}}
\item Dilate $c_{k}$ until it touches a mesh point $p_j$.
   Note that  $p_j$ might well be  $p_{k}$, $q_{k}$, or $t_{k}$, so, 
   $c_k$ might not actually expand at all during this step.
\item Grow the circle by moving its center away from
  $p_{j}$ along the ray $\overrightarrow{p_j\dot{c}_{k}}$,
  and maintaining the property that $p_{j}$ lies on the boundary
  of the circle,
  until it touches a mesh point $q_{j}$.
\item Continue to grow the circle, maintaining its contact
  with $p_{j}$ and $q_{j}$, moving its center away from
  the chord $\overline{p_j q_j}$,
  until it touches a vertex $t_j$.
\end{enumerate}
\vspace{-6pt}
The resulting circle $c'_j$ is a circumcircle of a Delaunay
  triangle $p_j q_j t_j$ at iteration $j$.
Moreover, $p_j q_j t_j$ is a poorly-shaped triangle 
  because its circumradius $r'_j$
  is at least $r_{k}$.
Thus, its center $\dot{c}'_j$ is a candidate at iteration $j$.
Note also that  $r'_j \geq r_{k} \geq 3r_i/4$.

Consider the triangle $p_j\dot{c}_{k}\dot{c}'_j$, 
  which is non-acute at vertex $\dot{c}_{k}$.
Let $x=|\dot{c}_{k}p_j|$ and $y=|\dot{c}_{k}\dot{c}'_j|$. 
Since $|\angle{\dot{c}'_j\dot{c}_{k}p_j}|$ is non-acute
  $(r'_j)^2 \ge x^2 + y^2$.
As $r'_{j}$ is the radius of a Delaunay triangle and $j \geq i$,
  $r'_{j} \leq r_{i}$.
Combining this fact with 
  $x \ge r_{k} > 3r_i/4$, we find $r'_j < x + r_i/4$.
So we can write, $(x+r_i/4)^2 \ge x^2 + y^2$.
By simplifying this inequality to  $x r_i/2+ r_i^2/16 \ge y^2$ and
  substituting $x \le r_i$, we derive $9r_i^2/16 \ge y^2$.
Hence, $y =||\dot{c}'_j -\dot{c}_{k}|| \le 3r_i/4$.

Because $c'_{j}$ is empty at the end of iteration $j$,
  we know $\dot {c'_{j}}$ was not chosen during iteration $j$.
Because the independent set of candidates that we select is maximal,
  there must be another circumcircle $c''_{j}$ chosen
  in iteration $j$
  that conflicts with $c'_{j}$.
By Lemma \ref{lem:conflict_reject_periodic},  
  the radius of $c''_j$ is at least one half of the radius of $c'_j$,
  and so is at least $|r'_j|/2 \geq 3r_i/8$.
Moreover, the radius of $c''_j$ is at most $r'_j$ and hence at most $r_{i}$.
So $||\dot{c}''_j - \dot{c}'_j|| \leq r_{i}$.
Hence,  $$ ||\dot{c}''_j -\dot{c}_{k}|| \leq 
  ||\dot{c}''_j -\dot{c}'_j|| + ||\dot{c}'_j - \dot{c}_{k}||
  \leq r_i + 3r_i/4 \leq 7r_i/4.$$

Let $\dot{C}''=\{ \dot{c}''_{i+1}, \dot{c}''_{i+2}, 
  \hdots, \dot{c}''_{j}, \hdots,   \dot{c}''_{k} \}$, and
  let $C''$ be the corresponding set of circumcircles.
As $\dot{c}''_{l}$ is inserted during round $l$ for each $l$,
  each circle $c''_{j} \in C''$ 
  is empty of all the centers $\dot{c}''_{l}$ for $l<j$.
So the centers in $\dot{C}''$ are pairwise at least $3r_i/8$ away
  from each other.
Thus, one can draw disjoint circles of radius $3 r_{i}/ 8$
  around each of these points.
The annulus containing these disjoint circles has area at most 
  $\pi(7/4 + 3/16)^2r_i^2 - \pi(3/4 - 3/16)^2 r_i^2$.
So, one can pack at most  
$$\Floor{\frac{\pi[(7/4 + 3/16)^2 - (3/4 - 3/16)^2]r_i^2}
                  {\pi(3/16)^2r_i^2}} = 97$$
 disjoint circles of radius $3 r_{i} / 16$ in this region.
This implies  $|C''| = k-i \leq 97$, a contradiction. 
\end{proof}

\begin{theorem}\label{thm:parallelchew_periodic}
Our parallel implementation of Chew's refinement algorithm 
  takes at most $\Ceil{98\log_{4/3}(L/s)}$ iterations.
\end{theorem}

\subsubsection{Parallel Computation of MIS}
\label{sec:MIS}

One can use the parallel maximal independent set algorithm 
  of Luby \cite{Luby86} to compute a parallel independent set of
  candidates for each iteration in $O (\log^{2} n)$ parallel time.
In this section, we will explain how we can exploit the geometric
  structure of the independence relation 
  to compute a maximal independent set in a constant parallel time.

We will make extensive use of the result of
  Lemma \ref{lem:conflict_reject_periodic} that
  two circumcircles are conflicting at iteration $j$ only if their
  radii are within a factor of 2 of each other.

\begin{lemma}\label{lem:constant}
At iteration $j$, if there are $n_{j}$ circumcircles,
  then a maximal independent set of candidates for Delaunay refinement
  can be computed in constant parallel time using $n_{j}$ processors.
\end{lemma}
\begin{proof}
Let $C^{j}_{h}$ be the set of circumcircles of radius more than
   $L/2^{h+1}$ and less than or equal to $L/2^{h}$, where
   $h$ ranges from $0$ to $\log (L/s_{j})$ and $s_{j}$ is the
   smallest circumradius at iteration $j$.
Note that a circumcircle in $C^{j}_{h}$ does not conflict with any 
  circumcircle in $C^{j}_{l}$ if $l > h + 1$.

To compute a maximal set of non-conflicting candidates, we 
  first in parallel find a maximal independent sets of circumcircles
  in $C^{j}_{h}$, independently for all even $h$.
We will show below that a maximal independent set of circumcircles in
  $C_{h}^{j}$ can be computed in constant time in parallel.
Let $I^{j}_{even}$ be the set of independent circumcircles computed.
Then in one parallel step, we can eliminate all conflicting
  circumcircles in $\cup_{h:odd} C^{j}_{h}$.
We then compute a maximal independent set for remaining circumcircles
  in $C_{h}^{j}$ for all odd $h$.
Let this set be $I^{j}_{odd}$.
Then $I^{j}_{even} \cup I^{j}_{odd}$ is a maximal 
  independent set of circumcircles
  for iteration $j$.

Note that all circumcircles in $C^{j}_{h}$ have radius between 
  $L/2^{h+1}$ and $L/2^{h}$.
If we divide the square containing all circumcenters into
  a $2^{h}$-by-$2^{h}$ grid, then
  any circumcenter that is conflict with 
  a circumcenter in the grid box $(x,y)$ must lie
  either in grid box $(x,y)$ or one of its eight grid neighbors.

We color grid boxes $(x,y)$ with color $(x\mod 3, y\mod 3)$.
We then cycle through the 9 color classes and, 
  by a method we will explain momentarily, 
  find a maximal
  independent set of the candidates in each 
  grid-box of the current color in parallel.
We then eliminate in parallel the conflicting circumcenters that are in the
  color classes that have not yet been processed.

Finally, we explain how to compute a maximal independent set among
  the candidates that lie in a given grid-box.
First notice that any maximal independent set of candidates in a grid-box
  can have at most a constant number of members, and hence
 a maximal independent set can be found by a constant number of
  parallel selection-elimination operations: choose a center that has not been
  eliminated, and in parallel eliminate any centers with which it
  conflicts.

In a parallel system that supports primitives such 
  fetch\_and\_add, test\_and\_set,
   or parallel scan, we can use such a primitive
  to select in constant time a candidate in a grid box.
The processor that holds this candidate becomes a ``leader''
  in that round and broadcasts its candidate so that the conflicting
  candidates can be eliminated.
With these primitives, our algorithm can be implemented in 
  parallel constant time.
However, if the parallel system does not support
  these primitives, then for each grid cell we can emulate
  parallel scan to select a leader in $O (\log n)$ time, where
  $n$ is the number of candidate centers in the cell.

In general, many grid cells are empty and there is no need to
  generate them at all. 
We can use hashing to select grid cells that are not empty.
The idea is very simple, each candidate center can compute
  the coordinates of its grid cell from the coordinates of its center
  and its radius.
We can hash grid cells using their coordinates and therefore,
  all candidate centers belonging to a grid cell can independently
  generate the hash identity of the cell.
We can then use parallel primitives discussed in the paragraph above
  to support the computation of a maximal independent set of
  candidates for all non-empty grid cells.
\end{proof}

\subsubsection{Parallelizing Ruppert's Refinement (PPS)}
\label{sec:parallelruppert}

In this section, we show that our parallelization of
  Ruppert's method for 
  periodic point sets in 2D takes $O (\log^{2} (L/s))$ iterations.
For simplicity,
   we give an analysis for the case $\beta_R = \sqrt{2}$, 
  although our analysis can be easily extended 
  to the case when $\beta _R = 1 + \epsilon$, 
  for any $\epsilon > 0$.
We recall that $\beta_{R}$ is the threshold of the ratio
  of the radius to shortest edge-length defining a poorly shaped triangle.
Thus, for $\beta_{R} = \sqrt{2}$,
  inserting the circumcenter of a poorly shaped triangle
  whose shortest edge is $h$ introduces new Delaunay
  edges of length at least $\sqrt{2}h$.

\begin{algorithm}
\caption{Parallel Ruppert's Refinement} 
\label{alg:ruppertpoint}
\begin{algorithmic}
\REQUIRE A periodic point set $P$ in $\RR^2$

\STATE  Let $T$ be the  Delaunay triangulation of $P$

\FOR{i=1 to $\lceil \log_{\sqrt{2}}(L/s) \rceil$}
\STATE Let $\dot{\CC}$ 
        be the set of all circumcenters of poorly-shaped
        triangles who are in class $\EE_i$
\WHILE{$\dot{\CC}$ is not empty}
\STATE Let $\II$ be a maximal independent subset of $\dot{\CC}$
\STATE Insert all the points in $\II$ in parallel
\STATE Update the Delaunay triangulation and $\dot{\CC}$ 
\ENDWHILE
\ENDFOR
\end{algorithmic}
\end{algorithm}

Let $s$ be the length of the shortest edge in the initial Delaunay 
  triangulation. 
At each iteration, we assign an edge to class $\EE_{i}$
  if its length is in $\IntervalCO{ \sqrt{2}^{i-1}s, \sqrt{2}^{i}s }$.
Similarly, we assign a Delaunay triangle to $\EE_{i}$
  if its shortest edge has length in 
  $\IntervalCO{ \sqrt{2}^{i-1}s, \sqrt{2}^{i}s }$.
There are at most $\lceil \log_{\sqrt{2}}(L/s) \rceil$ of such classes.
Using this definition, we can state and 
  analyze the Parallel Ruppert's Refinement Algorithm.

\begin{theorem}\label{thm:parallelruppert}
Given a periodic point set in 2D of diameter $L$, 
  the Parallel Ruppert's Refinement Algorithm
  takes $O(\log^{2}(L/s))$ iterations.
\end{theorem}

\begin{proof}
Lemmas \ref{lem:rupertConservation} and \ref{lem:upgrade} prove
  that after the $i$th iteration of the outer loop, 
  each Delaunay triangle touching an edge in 
  class $\EE _i$ will be well-shaped, and 
  successive iterations cannot degrade the shape of the Delaunay
  triangles touching that edge.
Lemma \ref{lem:techruppert} implies that during each iteration of the
  outer loop, the inner loop of the algorithm will execute at most
  $O(\log(L/s))$ times.
As the outer loop is executed $O(\log(L/s))$ times, the whole
  algorithm takes at most $O(\log^{2}(L/s))$ iterations.
\end{proof}

\begin{lemma}\label{lem:rupertConservation}
During the $i$th iteration of the outer loop of
  the Parallel Ruppert's Refinement Algorithm, 
  no Delaunay edges are added to or removed from
  class $\EE _i$.
\end{lemma}

\begin{lemma}
\label{lem:upgrade}
Suppose $e$ is an edge in $\EE_{j}$ where $j\leq i$. 
Then during the $i$th outer loop, 
  the radius-edge ratio of triangles containing $e$
  does not increase.
\end{lemma}

\begin{lemma} 
\label{lem:techruppert}
Let $e \in \EE_{i}$, and
 let $r_{l}$ be the radius of the larger of the two circumcircles
  containing $e$ at the end of the
  $l^{th}$ iteration of the inner loop during the $i$th iteration
  of the outer loop.
Then, at the end of 
  iteration $k = l+81$ of the inner loop,
  either (1) both Delaunay triangles containing $e$ are
  well-shaped, or (2) $r_{k} \le 3r_{l}/4$ where
  $r_{k}$ is the radius of the larger of the two circumcircles
  containing $e$ after iteration $k$.
\end{lemma}

\subsection{Input Domain: {PSLG}}\label{sec:pslg}

In this subsection, we extend our parallel algorithm for
  generating a Delaunay mesh from  a domain given by a 
  periodic point set to a domain defined by a planar
  straight-line graph.
Following Ruppert, we assume that the angle between two
  adjacent input segments is at least $\pi /2$.
A key step in Delaunay refinement for a domain specified by a PSLG 
  is to properly add points to the boundary segments so that 
  the Delaunay mesh is conforming to the boundary.
In our parallel algorithm, we make our mesh conform to the
  boundary in two steps:
First, we give, in Section \ref{sec:preprocessing}, 
    an $O (\log L/s)$ time parallel preprocessing algorithm
    to insert points to input segments
    so that the initial Delaunay mesh
    is conforming to the boundary and no diametral circle 
    intersects any other non-incident input features.
Second when a segment is encroached during parallel Delaunay refinement,
    we include its midpoint as candidate for insertion.

The preprocessing step might not be needed to implement
  our parallel algorithm: one could probably add points to the boundary
  as needed.
However, the preprocessing step simplifies
  our analysis in this Section by greatly reducing the number of cases
  in the analysis.

\subsubsection{A generic parallel algorithm (PSLG)}

After applying the preprocessing step, the initial 
  Delaunay triangulation is conforming to the boundary and 
  no diameter circle contains any point of the triangulation.
We will maintain this invariant in our algorithm.

In order to perform parallel refinement, as in Section 
  \ref{sec:genericParallelDelaunayRefinement}, 
  we need a rule of {\em independence} among
  candidates for refining boundary segments and 
  poorly shaped triangles.
We first recall the set of candidates for insertion
  defined in Section \ref{sec:DelaunayRefinement}.

Let $\BB$ be the set of circumcircles of poorly shaped triangles whose
   centers $\dot{\BB}$ encroach some boundary segments.
Let $\CC$ be the set of circumcircles of poorly shaped triangles whose
   centers $\dot{\CC}$ don't encroach any boundary segments.
Let $\DD$ is the set of diametral circles that are encroached by some
  centers in $\dot{\BB }$.
So, $\dot{\CC} \cup \dot{\DD} $ are candidate points for insertion.

We will still apply Definition \ref{def:conflict} to determine whether
  two circumcenters from $\dot{\CC} $ are independent.
Because the angle between two  adjacent input segments is at least $\pi /2$,
  after preprocessing, any two diametral circles from $\dot{\DD }$ are
  not overlapping.
Every two diametral centers from $\dot{\BB }$ are independent.

We will use the following definition of independence 
  between a diametral center in $\dot{\DD }$ and a 
  circumcenter in $\dot{\CC }$.
Note that because a circumcenter in $\dot{\CC}$ does not encroach any
  boundary segment, a diametral circle of $\DD$ does
  not contain any center in $\dot{\CC }$. 

\begin{definition}
\label{def:conflict_diametral}
A circumcenter  $\dot{c} \in \dot{\CC}$ 
  and a diametral center $\dot{d} \in \dot{\DD}$ 
  are {\em conflicting} if 
(i) $\dot{d}$ is inside $c$; and (ii)
  the radius of $c$ is smaller than $\sqrt{2}$ times the
  radius of $d$. 
Otherwise, $\dot{c}$ and $\dot{d}$ (also $c$ and $d$) are 
  {\em independent}. 
\end{definition}

This definition of independence is motivated by the following
  lemma first proved by Ruppert \cite{Ruppert93}. 

\begin{lemma}\label{lem:encroach-ratio}
If a circumcircle $c$ of radius $r_{c}$
  encroaches a diametral circle $d$ of radius $r_{d}$, then
  $r_{d} \geq r_{c}/\sqrt{2}$.  
\end{lemma}

\begin{algorithm}
\begin{algorithmic}
\caption{Generic Parallel Delaunay Refinement}
\label{alg:genericParallelDelaunayRefinementPSLG}

\REQUIRE A domain $\Omega$ given by a PSLG in $\RR^2$
\STATE Apply the parallel preprocessing algorithm of Section 
  \ref{sec:preprocessing}
\STATE Let $T$ be the initial Delaunay triangulation.
\STATE Compute $\dot{\BB \CC}$, an independent subset of  
       $\dot{\BB} \union \dot{\CC}$
\STATE Let $\dot{\DD}$ be the set of centers of diametral circles 
       encroached by the centers in $\dot{\BB \CC}$
\WHILE{$(\dot{\BB \CC} \cap \dot{\CC}) \union \dot{\DD}$ is not empty}
\STATE Let $\II$ be an independent subset of 
            $(\dot{\BB \CC} \cap \dot{\CC}) \union \dot{\DD}$
\STATE Insert all the points in $\II$ in parallel 
\STATE Update the Delaunay triangulation
\STATE Update $\dot{\BB}$, $\dot{\CC}$, $\dot{\BB \CC}$ and $\dot{\DD}$
\ENDWHILE
\end{algorithmic}
\end{algorithm}

The following theorem extends Theorem
  \ref{thm:sequential_parallel_periodic} 
  for domains given by PSLGs.

\begin{theorem}
\label{thm:sequential_parallel}
For a domain $\Omega $ specified by a PSLG, 
 suppose $M$ is a mesh produced by an execution of the parallel
 algorithm above.
Then $M$ can be obtained by some execution of 
  one of the sequential Delaunay refinement
  algorithms discussed in Section \ref{sec:DelaunayRefinement}.
\end{theorem}

\subsubsection{Parallelizing Chew's Refinement (PSLG)}
\label{sec:chew}
To parallelize Chew's algorithm for domain defined by a PSLG,
  we apply Algorithm \ref{alg:genericParallelDelaunayRefinementPSLG} 
  and use a maximal independent set of the candidates at each
  iteration.
In addition, because each pair of diametral centers in $\dot{\DD }$ is independent,
  we include all centers $\dot{\DD }$ in the independent set.
The parallel algorithm of Section \ref{sec:MIS} can be used to
  construct the maximal independent set.

\begin{theorem}
\label{thm:parallelchew}
Our parallel implementation of Chew's refinement algorithm 
  takes $O(\log (L/s))$ iterations 
  for a domain given by a PSLG, where  $L$ is the diameter of the
  domain and $s$ is smallest local feature size.
\end{theorem}

\subsubsection{Parallel Preprocessing}
\label{sec:preprocessing}

In the algorithm and proof presented in the last subsection,
  we assume that the boundary of the domain has been preprocessed
  to satisfy the following property.

\begin{definition}[Strongly Conforming]
\label{def:weakly}
A domain $\Omega$ specified by a PSLG is {\em strongly
  conforming} if no diametral circle contains
  any vertex or intersects any other non-incident input features.
\end{definition}

Clearly, if $\Omega $ is strongly conforming, 
  then the Delaunay triangulation of the 
  vertices of $\Omega $ is conforming to $\Omega $.

We will use the following parallel method to preprocess
  a domain $\Omega $ to make it strongly conforming.
This method repeatedly adds midpoints to boundary segments whose
  diametral circles intersect non-incident input features.

\begin{algorithm}
\caption{Parallel Boundary Preprocessing} 
\begin{algorithmic}
\REQUIRE A PSLG domain $\Omega$ in $\RR^2$
\STATE  Let $\GG$ be the set of segments in $\Omega$ whose diametral
  circles intersect non-incident input features.
\WHILE{$\GG$ is not empty}
\STATE Split all the segments in $\GG$ in parallel
      by midpoint insertion and update $\GG$.
\ENDWHILE
\end{algorithmic}
\end{algorithm}

\begin{lemma}\label{lem:preprocessing}
Parallel Boundary Preprocessing terminates in $O(\log(L/\ss))$
  iterations. 
\end{lemma}

In the scheme above, we can grow a quadtree 
  level by level to support the query of
  whether the diametral circle of a segment intersects another
  non-incident feature.
The number of levels of the quadtree
   that we need to grow is at most $\log (L/s)$.
As shown in \cite{BernEG94,BernET99}, one can use
  balanced quadtree to approximate local feature size function of
  $\Omega $ to within a constant factor.
Therefore, using a balanced quadtree as \cite{BernEG94,BernET99},
  we can preprocess the domain in $\log (L/s)$ parallel time so
  that the preprocessed domain is {\em strongly feature conforming}
  as defined below.

\begin{definition}[Strongly Feature Conforming]
\label{def:strongly} Let $\alpha >2 $ be a constant.
A domain $\Omega$ specified by a PSLG is {\em strongly
  feature conforming} with parameter $\alpha $
  if it is strongly conforming,
  and in addition, the length of each segment is no more than $\alpha
  $ times the local feature size of its midpoint.
\end{definition}

In the next subsection, we will present a 
  parallel implementation of Ruppert's algorithm for domains that
  are strongly feature conforming and show that it
  terminates in $O (\log^{2} (L/s))$ iterations.

We use the following lemma to show that the size optimality of our
  results are not affected much by the preprocessing.

\begin{lemma}\label{lem:lfs_ratio}
Let $\Omega$ and $\Omega'$ denote the input before and after 
   preprocessing, respectively. 
Then, for any point $x$ in these domains, 
  $\lfs_{\Omega}(x)/3 \le \lfs_{\Omega'}(x) \le \lfs_{\Omega}(x)$.
\end{lemma}

\subsubsection{Parallelizing Ruppert's Refinement (PSLG)}

In this section, we show that our parallelization of
  Ruppert's method for a domain given by a PSLG takes
  $O (\log^{2} (L/s))$ iterations.
Again, for simplicity, we will only
   give an analysis for the case when $\beta_R = \sqrt{2}$.

The parallel algorithm follows basic steps of 
 the parallel Ruppert's Refinement presented earlier in 
 Section \ref{sec:parallelruppert}.
But first, we apply the parallel preprocessing 
  algorithm of Section \ref{sec:preprocessing} so that the 
  preprocessed domain is strongly feature conforming.
So below we can assume that $\Omega $ is strongly 
  conforming.

Let $s$ be smallest local feature of $\Omega $.
At each iteration, we 
  assign an edge to class $\EE_{i}$
  if its length is in $\IntervalCO{ \sqrt{2}^{i-1}s, \sqrt{2}^{i}s }$.
Similarly, we assign a Delaunay triangle to $\EE_{i}$
  if its shortest edge has length in 
  $\IntervalCO{ \sqrt{2}^{i-1}s, \sqrt{2}^{i}s }$.
There are at most $\lceil \log_{\sqrt{2}}(L/s) \rceil$ of such classes.

\begin{algorithm}
\caption{Parallel Ruppert's Refinement} 
\label{alg:ruppert}
\begin{algorithmic}
\REQUIRE A domain $\Omega$ given by a PSLG that is strongly feature 
  conforming.
\STATE Let $T$ be the initial Delaunay triangulation.

\FOR{i=1 to $\lceil \log_{\sqrt{2}}(L/s) \rceil$}
\STATE Let $\dot{\BB}$  
       be encroaching candidate circumcenters and $\dot {\CC }$ be the
       non-encroaching candidate circumcenters whose 
       triangles is in class $\EE_i$.
\STATE
Compute $\dot{\BB \CC}$, an independent set of  
       $\dot{\BB} \union \dot{\CC}$.
\STATE Let $\dot{\DD}$ be the set of centers of diametral circles 
       encroached by the centers in $\dot{\BB \CC}$
\WHILE{$(\BB \CC \cap \CC) \union \DD$ is not empty}

\STATE Let $\II$ be an maximal independent subset of 
            $(\dot{\BB \CC} \cap \dot{\CC}) \union \dot{\DD}$
\STATE Insert all the points in $\II$ in parallel
\STATE Update the Delaunay triangulation
\STATE Update $\dot{\BB}$, $\dot{\CC}$, $\dot{\BB \CC}$ and $\dot{\DD}$
\ENDWHILE
\ENDFOR
\end{algorithmic}
\end{algorithm}

\begin{theorem}\label{thm:parallelruppertPSLG}
Given a domain specified by a PSLG,
  the Parallel Ruppert's Refinement Algorithm
  takes $O(\log^{2}(L/s))$ iterations.
\end{theorem}

The proof of Theorem \ref{thm:parallelruppertPSLG} is essentially 
  the same as the proof of Theorem \ref{thm:parallelruppert} where
  we need to address the following two issues.

\vspace{-6pt}
\begin{enumerate}
\setlength{\itemsep}{-2pt}
\item The center of a circumcircle could potentially 
      encroach a boundary segment whose length is much larger than 
      that the circumradius.
\item The insertion of a midpoint on the boundary could potentially
      introduce smaller edges.
\end{enumerate}
\vspace{-6pt}

To address the first issue, we apply parallel processing algorithm 
 of Section \ref{sec:preprocessing} and hence assume $\Omega $ 
  is strongly feature conforming.
Hence if a circumcenter encroaches a boundary
  segment, the circumradius and the length of the segment are
  with a constant factor of each other.
In addition, because each boundary segment can only be
  split at most a constant times in the refinement, 
  it can not introduce smaller edges too many times.

\section{3D Delaunay Refinement}
\label{sec:3d}

A 3D domain is specified by a PLC (see Section \ref{sec:input}).
In this section, we assume 
  that the angle between any two intersecting elements,
  when one is not contained in the other, is at least $90^\circ$.
There are three kinds of spheres associated with a 3D Delaunay mesh
  that we are interested:
  the circumspheres, the diametral spheres, and the equatorial sphere
  given below.

\begin{definition}
The {\em equatorial sphere} of a triangle in 3D
  is the smallest sphere that passes
  through its vertices.
A triangular subfacet of a PLC is {\em encroached} if the
  equatorial sphere is not empty.
\end{definition}

Chew's algorithm extends naturally to 3D.
In \cite{Shewchuk98}, Shewchuk developed a 3D
  extension of Ruppert's algorithm.
In Shewchuk's refinement, given below, a tetrahedron is {\em bad} 
  if the ratio of its circumradius to its shortest edge, referred as
  the radius-edge ratio, is more than a pre-specified 
  constant $\beta_S \ge 2$.

\begin{algorithm}
\caption{3D Delaunay Refinement} 
\begin{algorithmic}
\REQUIRE  A PLC domain $\Omega$ in $\RR^3$

\STATE Compute $T$, the Delaunay triangulation of the points of $\Omega$
\STATE Let $\dot{C}$ be the set of non-encroaching circumcenters of the 
      bad tetrahedra
\STATE Let $\dot{D}$ be the set of non-encroaching equatorial centers of 
      the encroached triangular subfacets
\STATE Let $\dot{E}$ be the set of diametral centers of the 
      encroached subsegments.

\WHILE{there is center $a$ in $\dot{C} \union \dot{D} \union \dot{E}$ is not empty}
\STATE Insert $a$ and update the Delaunay triangulation
\STATE Update $\dot{C}$, $\dot{D}$, and $\dot{E}$ 
\ENDWHILE
\end{algorithmic}
\end{algorithm}

\subsection{Parallel 3D Delaunay Refinement}
\label{sec:parallel3D}

In this subsection, we show that our results for a domain given by a 
  periodic point set can be extended from two dimensions 
  to three dimensions to parallelize both Chew's and Shewchuk's algorithm.
So far, we have not completed the analysis for domains specified by PLCs, 
  although we think similar results can be obtained.

The following is a parallel Delaunay refinement algorithm  for domains
  specified by 3D periodic point sets.

\begin{algorithm}
\caption{Generic 3D Parallel Delaunay Refinement} 
\label{alg:genericParallelDelaunayRefinement3D}
\begin{algorithmic}
\REQUIRE A periodic point set $P$ in $\RR^3$

\STATE Let $T$ be the Delaunay triangulation 
      of $P$ 
\STATE Compute  $\dot{\CC}$, the set of circumcenters of bad tetrahedra
in $T$
\WHILE{$\dot{\CC}$ is not empty}
\STATE Let $\II$ be an independent subset of $\dot{\CC}$
\STATE Insert all the points in $\II$ in parallel 
\STATE Update $\dot{\CC}$
\ENDWHILE
\end{algorithmic}
\end{algorithm}

To parallelize Chew's 3D refinement, we use a {\em maximal} 
  independent set of  candidate
  centers in Algorithm \ref{alg:genericParallelDelaunayRefinement3D}.
With almost the same proof as we have presented in Section 
  \ref{sec:chew_periodic}, we can show that the number of iterations
   needed is $1076\log (L/s)$.

\begin{theorem}
\label{thm:sequential_parallel_periodic_3D}
Suppose $M$ is a mesh produced by an execution 
  of the 3D Generic Parallel Delaunay Refinement algorithm.
Then $M$ can be obtained by some execution of 
  the sequential Delaunay refinement algorithm.
\end{theorem}

\subsubsection{Parallelizing Shewchuk's Refinement}

We will present our analysis 
  for the case when $\beta_R = \sqrt{2}$,  
  although our analysis can be easily extended 
  to the case when $\beta _R = 1 + \epsilon$, 
  for any $\epsilon > 0$.
Thus, for $\beta_{R} = \sqrt{2}$,
  inserting the circumcenter of a poorly shaped triangle
  whose shortest edge is $h$ introduces new Delaunay
  edges of length at least $\sqrt{2}h$.

Let $s$ be the length of the shortest edge in the initial Delaunay 
  triangulation. 
At each iteration, we 
  assign an edge to class $\EE_{i}$
  if its length is in $\IntervalCO{ \sqrt{2}^{i-1}s, \sqrt{2}^{i}s }$.
Similarly, we assign a Delaunay tetrahedra to $\EE_{i}$
  if its shortest edge has length in 
  $\IntervalCO{ \sqrt{2}^{i-1}s, \sqrt{2}^{i}s }$.
There are at most $\lceil \log_{\sqrt{2}}(L/s) \rceil$ of such classes.

Our parallel implementation of Shewchuk's algorithm is 
  analogous to our parallel implementation of Ruppert's
  algorithm.
In addition, our proof in 3D is also analogous to the proof in 2D.

\begin{theorem}
\label{thm:parallelShewchuk}
For a given periodic point set $P$ in $\RR^3$ of diameter at most $L$, 
  if the length of the shortest edge in the mesh generated by
  Shewchuk's refinement is $s$, then parallel Shewchuk refinement takes
  $O(\log^{2} (L/s))$ iterations to generate a
  bounded radius-edge ratio mesh.
\end{theorem}

\section{Discussion}
\label{sec:discussion}

Polylogarithmic upper bounds on the number of parallel iterations 
  presented in Sections \ref{sec:parallelDelaunayRefinement} 
  and \ref{sec:parallel3D} constitutes the  
  main component of the analyses of our parallel algorithms.
At each iteration, our algorithms perform two main operations:
  i) compute a maximal independent set of points for parallel insertion;
 ii) update the Delaunay triangulation inserting all these points.
For the first one, we proposed a new constant time parallel algorithm. 
For the second, we suggested to use an existing logarithmic time 
parallel Delaunay triangulation algorithm.
These immediately imply polylogarithmic total time complexity for our
  parallel Delaunay refinement algorithms.

We opted for simplicity in our analyses. So, the constants in lemmas 
 \ref{lem:chew2d_constant_iterations_periodic} and
 \ref{lem:techruppert} are probably not optimal 
  and likely to be much smaller in practice than 98 and 81.

The 3D extension of Chew's and
 Shewchuk's  algorithms do not always guarantee
  that the resulting mesh has an aspect-ratio bounded by a constant.
However, they both guarantee a constant bound on 
  the ratio of the circumradius to 
  the length of the shortest edge (the radius-edge ratio) of any tetrahedra 
  in the final mesh. 
So, the meshes these two algorithms
  generate might potentially contains slivers, which are 
  elements with close to zero aspect-ratio but with a constant
  radius-edge ratio. 
Several quality enhancing and guaranteeing meshing algorithms
  \cite{ChengDEFT99, Chew97, EdelsbrunnerLMSTTUW00, LiTeng01} 
  have been developed recently.
Cheng {\em et al.}\ \cite{ChengDEFT99}
  and Edelsbrunner {\em et al.}\ \cite{EdelsbrunnerLMSTTUW00}
  have already given parallel complexity of their sliver removal
  algorithms.
Our framework can be used to analyze parallel complexity of the 
  other two algorithms, by Chew \cite{Chew97} and Li and Teng \cite{LiTeng01}.

We conclude the paper with two conjectures.
\vspace{-6pt}
\begin{itemize}
\setlength{\itemsep}{-2pt}
\item There is a parallel implementation of Ruppert's \cite{Ruppert93}
  and Shewchuk's \cite{Shewchuk98} algorithm that runs in
           $O(\log (L/s))$ iterations.
\item There is a parallel Ruppert's \cite{Ruppert93} 
  and Shewchuk's \cite{Shewchuk98} algorithm that runs in
  $O(\log n)$ time where $n$ is the input complexity.
Notice that Bern {\em et al.}\ \cite{BernET99} showed that the
  quadtree algorithm can be implemented in $O(\log n)$ time with $K$
  processors. 
\end{itemize}
\vspace{-6pt}
We would also like to see results that establish 
  the parallel complexity of other mesh generation algorithms
  such as sink insertion \cite{EdelsbrunnerG01}.

\section{Acknowledgments}\label{sec:ack}
We thank Jeff Erickson, Sariel Har-Peled and Dafna Talmor 
  for helpful conversations and comments on the paper.
We also thank Jonathan Shewchuk for helpful emails about
  boundary assumptions and boundary preprocessing
   for Ruppert's algorithm.

{\footnotesize
\bibliographystyle{newabuser}
\bibliography{parDelRef_short}
}

\end{document}